\documentclass[a4paper, 12pt]{article}
\usepackage{epsfig}
\usepackage{graphicx}
\usepackage{amsmath}
 \usepackage{rotating}
\begin{document}
\title{Criticality and the Onset of Ordering in the Standard Vicsek Model}
\author{Gabriel Baglietto$^{1,2}$, Ezequiel V. Albano$^{1,3}$ 
and Juli\'an Candia$^1$\\
$^1${\small\it Instituto de F\'{\i}sica de L\'{\i}quidos y Sistemas 
Biol\'ogicos (CCT-CONICET-La Plata, UNLP),}\\{\small\it 59 Nro 789, 1900 La Plata, Argentina}\\
$^2${\small\it Facultad de Ingenier\'{i}a (UNLP), La Plata, Argentina,}\\
$^3${\small\it Departamento de F\'{\i}sica, 
Facultad de Ciencias Exactas (UNLP),}\\ 
{\small\it La Plata, Argentina}}
\maketitle

\begin{abstract}
Experimental observations of animal collective behavior have shown stunning evidence for the emergence 
of large-scale cooperative phenomena resembling phase transitions in physical systems. Indeed, quantitative 
studies have found scale-free correlations and critical behavior consistent with the occurrence of continuous, 
second-order phase transitions. The Standard Vicsek Model (SVM), a minimal model of self-propelled particles 
in which their tendency to align with each other competes with perturbations controlled by a noise term, appears 
to capture the essential ingredients of critical flocking phenomena. In this paper, we review 
recent finite-size scaling and dynamical studies of the SVM, which present a 
full characterization of the continuous phase transition through dynamical and critical exponents. 
We also present 
a complex network analysis of SVM flocks and discuss the onset of ordering in connection with XY-like spin models. 
\end{abstract}

\section{Introduction}

Nature offers abundant manifestations of collective motion phenomena in self-propelled living systems at all scales, from biomolecular micromotors, migrating 
cells and growing bacteria colonies, to insect swarms, fish schools, bird flocks, mammal herds and even human crowds. Instead of focusing on the 
specific details that make each of these biological systems unique, 
statistical physicists have been studying the general patterns of biological collective motion, 
aiming to identify the general laws and underlying principles that may govern their behavior~\cite{vics10}.

From this perspective, one important question to address is the onset of ordered macroscopic phases, i.e. the way in which individuals 
having short-range interactions are capable of self-organizing into large-scale cooperative patterns in the absence of leaders or other ordering cues from the environment. By analogy with large molecular systems, flocking and swarming phenomena can be associated 
with phase transitions that depend on a few parameters that characterize the macroscopic states, such as the density of individuals and the flock size. 
For instance, Buhl et al.~\cite{buhl06} investigated the collective motion of locusts, which display a density-driven transition 
from disordered movement of individuals within the group to highly aligned collective motion. In analogy with second-order (or continuous) phase transitions 
in physical phenomena such as critical opalescence, in which liquids appear opaque due to density fluctuations at all wavelengths of visible light, the onset of a coordinated alignment in the motion of marching locusts was found to display second-order behavior. Moreover, a critical locust density for the occurrence of the continuous phase transition was identified.

Familiar to most of us is also the rich variety of collective motion phenomena displayed by flocks of birds, where highly correlated long-range ordering 
effects are readily apparent (see e.g.~\cite{bbc09} for some stunning footage of complex flocking patterns and other kinds of swarm behavior). 
Indeed, some studies have quantitatively characterized the formation of ordered bird flocks in terms of continuous phase transitions. For instance, 
Cavagna et al.~\cite{cava10} obtained high resolution spatial data of thousands of starlings using stereo imaging in order to calculate 
the response of a large flock to external perturbations. This study showed that behavioral correlations are scale-free, a signature of critical states in second-order phase transitions. Correlation functions revealing the pairwise interaction between homing pigeons have recently been investigated by means of high-resolution lightweight GPS devices as well~\cite{nagy10}. 

On the theoretical side, Vicsek et al.~\cite{vics95} proposed a minimal model to study the onset of order in systems of self-driven individuals, 
which was later followed by other investigations by means of agent-based modeling~\cite{czir99,alda07}, the Newtonian 
force-equation approach~\cite{mikh99,erdm05}, and the hydrodynamic approximation~\cite{tone95,tone05}. 
This so-called Standard Vicsek Model (SVM),
assumes that neighboring individuals tend to align 
their direction of movement when they are placed within a certain interaction range. This alignment rule, which would trivially lead to fully ordered collective motion, is complemented by a second one that introduces noise in the communications among individuals. 
The SVM assumes that each individual can assess exactly the direction of its neighbors, but it incurs an error when  
adjusting its own direction of motion to match that of its neighbors' average. This type of noise, sometimes identified as {\it Angular Noise}, 
was shown to drive the system through a second-order transition between the ordered phase of collective motion and a disordered phase. 
Although several variations to the SVM have later been considered in the literature, such as different noise types, models without alignment rule, adhesion 
between neighbor individuals, bipolar particles, etc (see~\cite{vics10} for a review), here we will chiefly focus on the original SVM as formulated by Vicsek et al. in their seminal paper~\cite{vics95}.

In this context, the aim of this work is to review recent findings for the SVM that confirm the continuous nature of the order-disorder phase transition 
driven by the noise amplitude, as well as to present a novel complex network analysis that provides new insight into the onset of ordering. 
The rest of this paper is organized as follows. In Section 2, we present the definition of the SVM. Section 3 reviews recent finite-size scaling and dynamical studies of the SVM, leading to a 
fully quantitative characterization of the continuous phase transition through dynamical and critical exponents.  
In Section 4, we present 
a complex network analysis of SVM flocks, we study their topology, and we discuss the onset of ordering in connection with XY-like spin models. Finally, our Conclusions appear in Section 5. 

\section{The Standard Vicsek Model (SVM)}
 
The SVM~\cite{vics95} consists of a fixed number of interacting particles, $N$, which are moving on a plane. 
In computer simulations, that plane is represented by a square of side $L$ with periodic boundary conditions~\cite{vics95,bagl09a}. 
The particles move off-lattice with constant and common speed $v_0\equiv |\vec{v}|$.
Each particle interacts locally adopting the direction of motion of the subsystem of neighboring particles (within an interaction circle of radius $R_0$ centered in the considered particle), 
which is then perturbed by the presence of noise. Since the interaction radius is the 
same for all particles, we adopt the interaction radius as the unit of length 
throughout, i.e. $R_0\equiv 1$. 

The evaluation of the average angle of motion of each individual's neighbors at time $t$, $\theta_j^t$, 
is affected by a noise term. Hence, the updated direction of motion for the $i-$th particle, $\theta_i^{t+1}$, is given by  
\begin{equation}
\theta_i^{t+1}=Arg\left[\sum_{\langle i,j\rangle}e^{i\theta_j^t}\right]+\eta\xi_i^t\ ,
\label{anterm}
\end{equation}
\noindent where $\eta$ is the noise amplitude, the summation is carried over all particles within the 
interaction circle centered at the $i-$th particle, and $\xi_i^t$ is a realization of a $\delta$-correlated white noise uniformly 
distributed between $-\pi$ and $\pi$. The noise term can be thought of as due to the error committed by the individual when trying
to adjust its direction of motion to the averaged direction of motion of its neighbors.

The model dynamics is usually implemented by adopting the so-called {\it backward update rule}: after the 
position and orientation (velocity) of all particles are determined at time $t$, the position of the particles at time $t+1$ is updated 
according to
\begin{equation}
\vec{x_i}^{t+1}=\vec{x_i}^t+\vec{v_i}^t\ ,
\label{BU}
\end{equation}
\noindent which is then followed by the update of all velocities at time $t+1$ according to Eq.(\ref{anterm}). 
However, different updating schemes implemented in the literature may lead to spurious effects in the simulations~\cite{bagl09b}. 

The parameters of the model are the number of particles $N$, the dimension of the displacement space $d$, the linear size of the displacement space $L$, the interaction radius $R_0$, 
the particle density $\rho$, the particle speed $v_0$, and the noise amplitude $\eta$. Notice, however, that $\rho=N/L^d$, 
where $d=2$ in the standard case, and that $R_0\equiv 1$ can be chosen as the unit of length. Moreover, $v_0^{-1}$ merely plays 
the role of a ``thermalization parameter" that 
measures how many times, on average, two neighbors check out each other's positions while they remain at a distance within the unitary interaction radius. Therefore, 
relative large values of speed (typically $v_0\geq 0.3$) correspond to the low-thermalization regime, characterized by highly anisotropic diffusion and the manifestation of  simulation artifacts in the form of directionally quantized density waves (see Ref.~\cite{nagy07} for details). The relevant region of parameter space addressed in this paper is the high-thermalization regime for $v_0\leq 0.1$ that guarantees isotropic diffusion. Within this regime, the choice of $v_0$ has only mild effects on the 
order parameter probability distributions and it does not affect the nature of the phase transition~\cite{bagl09a}.  

\section{Critical Behavior of the SVM: Finite-Size and Dynamic Scaling}   

The SVM exhibits a far-from-equilibrium continuous phase transition between ordered states of motion at 
low noise levels and disordered motion at high noise levels. The 
natural order parameter is given by the absolute value of the 
normalized mean velocity of the system ~\cite{vics95,bagl08}:
\begin{equation}
\varphi = \frac{1}{N v_0}\vert\sum_{i=1}^{N}\vec{v_i}\vert,
\label{orpa}
\end{equation}
\noindent where $\varphi$ is close to zero in the disordered phase and close to unity in the ordered phase. In order to anlyze the critical nature of the phase transition exhibited by the SVM,
we have studied behavior of the order parameter  $\varphi$ and other quantities related to the probability distribution of 
$\varphi$, such as the susceptibility ($\chi =\frac{N}{\rho}[\langle\varphi ^2\rangle-\langle\varphi \rangle^2 ]$) and the Binder 
cumulant ($U=1-\frac{\langle\phi^4\rangle}{3\langle\phi^2\rangle^2}$). The phase transition has been investigated  
by means of two independent approaches, namely a finite-size scaling analysis and a short-time critical 
dynamic technique~\cite{bagl08,RPP}.

Let us briefly mention that the second-order nature of the SVM far-from-equilibrium phase transition has been a matter of some debate. In fact, early simulations by Vicsek's group, which were consistent with a continuous second-order transition~\cite{vics95}, were later challenged by Gr\'egoire and Chat\'e~\cite{greg04}, who claimed that the transition was discontinuous, i.e. of first order. The controversy was further stimulated by subsequent papers of Vicsek et al.~\cite{nagy07,gonc08}, 
Aldana et al.~\cite{alda07,alda03,pime08,alda09}, Dossetti et al.~\cite{doss09}, and Baglietto and Albano~\cite{bagl09a,bagl09b,bagl08}, supporting the critical nature of the transition, which were in conflict with additional results 
published by Chat\'e and collaborators~\cite{chat07,chat08,chat08b}. 
In the remainder of this work, we assume that the SVM (using the so-called Angular Noise type) undergoes a second-order phase transition. 
For further details, see Refs.~\cite{bagl09b,nagy07,alda09}. 

\begin{figure}[t!]
\centerline{\epsfysize=3.3in\epsfbox{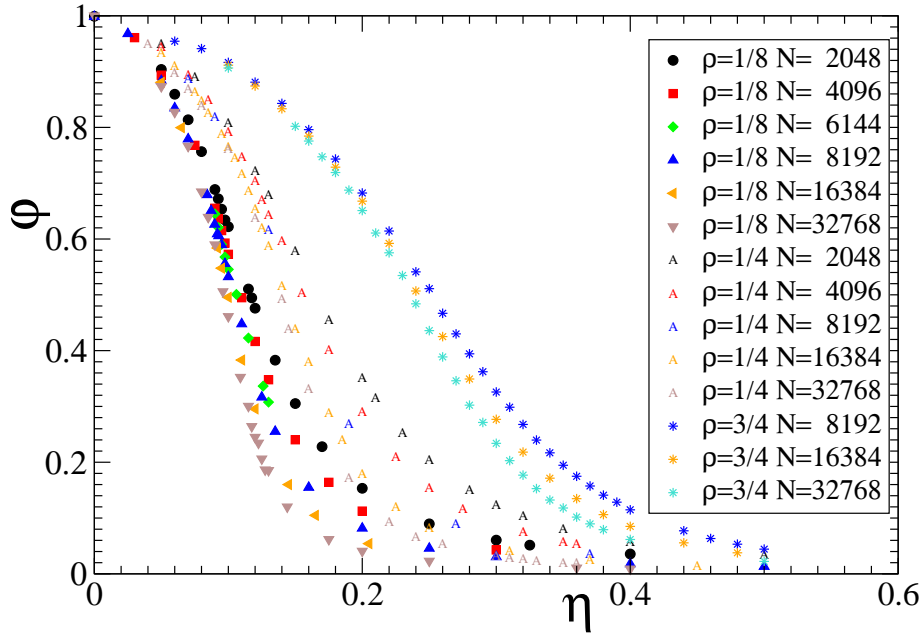}}
\caption{Plots of the order parameter ( $\varphi$) versus the noise amplitude ($\eta $).
Results obtained within the stationary regime for samples of different size 
and by varying the number of individuals, as listed in the figure~\cite{bagl08}.}
\label{op_brutos}
\end{figure}

\begin{figure}[t!]
\centerline{\epsfysize=3.3in\epsfbox{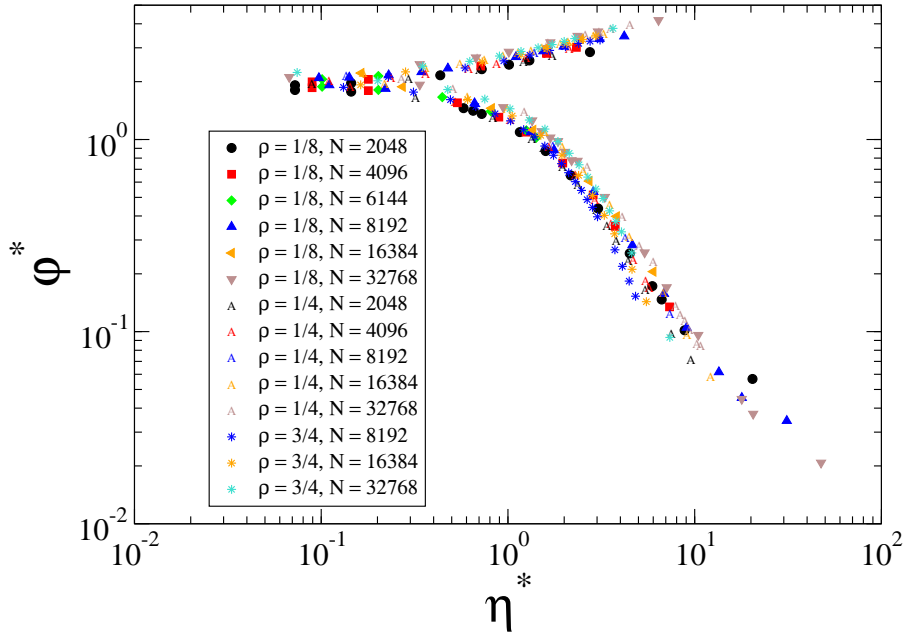}}
\caption{Finite-size scaling analysis of data obtained under stationary conditions 
for different densities and numbers of individuals, showing log-log plots of 
the rescaled order parameter $\varphi^{*} = \varphi(L)L^{\beta/\nu}$ 
versus the density-rescaled noise 
$\eta^{*} = \frac{\vert \eta(\rho) - \eta_c(\rho)\vert}{\sqrt{\rho}}L^{1/\nu}$ ~\cite{bagl08}.
Here, the leftmost (rightmost) points correspond to data measured closest to (farthest from) the critical point.}
\label{op_colapso}
\end{figure}

While critical systems are characterized by the divergence of correlation length and time scales, 
computational simulations are limited to systems of finite extent and finite time observation intervals. 
In systems of finite size $L$ (where, in our case, $L = \sqrt{\frac{N}{\rho}}$) the correlation length cannot grow beyond $L$. 
Instead of the theoretically expected power-laws at criticality 
(e.g. $\varphi \sim |\tau|^{\beta}$ and $\chi \sim |\tau|^{-\gamma}$, 
where $\tau$ represents the distance to the critical point and $\beta$ and $\gamma$ are critical exponents), 
for finite-size samples of critical systems, stationary quantities must obey scaling laws of the form:

\begin{equation}
\varphi (\tau,L)=L^{-\beta /\nu }\widetilde{\varphi }(\tau L^{1/\nu }),
\label{fss_op}
\end{equation}

\begin{equation}
\chi (\tau,L)=L^{\gamma /\nu }\widetilde{\chi }(\tau L^{1/\nu }),
\label{chi_finito}
\end{equation}
\noindent where $\widetilde{\varphi}$ and $\widetilde{\chi}$ are scaling functions, and 
$\nu $ is the correlation length critical exponent. 
In the case of the SVM, a very good agreement between simulation results and Eqs.(\ref{fss_op})-(\ref{chi_finito}) 
are obtained. Fig.~\ref{op_brutos} shows the behavior of $\varphi$ vs $\eta$ for different sizes and densities,
while Fig.~\ref{op_colapso} displays the collapse obtained by applying Eq.(\ref{fss_op}) to a known scaling law for the 
densities~\cite{bagl08}. The values of $\beta$ and $\nu$ used in the simulations are listed in Table~\ref{TableI}.

On the other hand, when a critical system, initially placed either in a ground state or in the disordered phase,
is suddenly brought to the critical point, it is also possible to measure its critical 
dynamic evolution. In this way, one can obtain not only the static 
exponents ($\beta$, $\nu$, and $\gamma$) but also the dynamic exponent $z$ that
governs the time-dependent growth of the correlation length~\cite{RPP}.  
Table~\ref{TableI} summarizes the exponents obtained for the SVM by means of stationary and dynamic measurements~\cite{bagl08}. 

\begin{table}[tbp]
\label{TableI}
\caption{Exponents of the SVM as obtained by using different methods. 
The acronyms S, RD, and DDC refer to stationary measurements, relaxation dynamics from ordered states, 
and dynamic measurements starting from disordered configurations, 
respectively~\cite{bagl08}.  
}
\vspace{1.0cm}
{\centering\resizebox{14.2cm}{!} { \Large\Large\Large\Large\begin{tabular}{|c|c|c|c|c|c|c|c|c|c|c|}
\hline 
& $\gamma/\nu$ & $z$ & $\gamma/(\nu z)$ & $\beta/(\nu z)$ & 
$1/(\nu z)$ & $\beta$ & $\nu$ & $\gamma$ & $\beta/\nu$\\ 
\hline
S & 1.45(2) & - & - & - & - & 0.45(3) & 1.6(3) & 2.3(4) & 0.275(5) \\
\hline 
RD & - & 1.27(2) & - & 0.25(2) & 0.6(1) & 0.42(4) & 1.3(3) & - & 0.32(3) \\ 
\hline
DDC & - & - & 1.12(3) & - & - & - & - & - & - \\ 
\hline
DDC+RD & 1.43(3) & - & - & - & - & - & - & 1.87(4) & - \\ 
\hline
S+RD & - & - & 1.13(3) & 0.22(2) & 0.5(1) & $0.45(7)^{*}$ & - & 1.89(4) & - \\ 
\hline
S+DDC & - & 1.29(3) & - & - & 0.5(1) & - & - & - & -\\ 
\hline
\end{tabular}\par}}
\end{table}
\vspace{1.0cm}

Considering statistical errors, the obtained exponents satisfy the hyperscaling relationship 
($d\nu -2\beta =\gamma$), which is firmly established for critical systems in equilibrium
but less documented for the case of nonequilibrium critical systems (see e.g. \cite{canalb}). 
This finding is a strong 
indication that the SVM undergoes  a continuous nonequilibrium phase transition.
It worth mentioning  that the set of critical exponents found for the SVM constitutes 
a new universality class. Also, the agreement between results obtained from stationary and 
dynamic measurements provides strong support to the SVM's critical nature. 

\section{The Onset of Orientation Ordering in SVM Flocks}

The SVM, which describes a far-from-equilibrium phenomenon, has often been compared with 
another archetypical case in the study of critical phenomena but under equilibrium conditions, namely the XY model~\cite{binn92,chai95}. 
In the XY model, nearest-neighbor interacting spins may adopt any possible orientation depending on the strength of the
interactions and the temperature. By considering the SVM as a model of interacting spins that can undergo displacements in the direction of the
spin, the basic difference between both models is precisely due to those displacements. In fact, 
other relevant symmetries for the study of phase transitions, such as the dimensionality of the space, the nature of the order parameter, 
and the range of the interactions, are the same in both models. On the other hand, it is well known that, according 
to the Mermin-Wagner theorem~\cite{merm66,merm67}, the XY model,
as well as all other equilibrium systems in $2D$ with short-range interactions and $O(2)$ symmetry defined on translationally-invariant 
substrates, cannot display ordered phases. 
Therefore, the onset of ordering in the SVM in $2D$ is quite intriguing. What is, then, the essential ingredient that generates the ordered phase 
of SVM flocks, despite Mermin-Wagner?
Naturally, the possible answers lie in the conditions required by the theorem to apply, which are not met by the SVM, namely:
\begin{itemize}
\item the SVM is non-equilibrium (e.g. particle interactions do not conserve momentum)
\item SVM ÒspinsÓ move and originate effective long-range interactions (flocks merge and dismember, carrying information across the system)  
\item The SVM ÒsubstrateÓ is not translationally invariant (particle positions are distributed inhomogeneously in space)
\end{itemize}

It is worth mentioning that the XY model defined on $1D$ small-world networks (generated by the addition of a small number 
of random long-range ``shortcuts" on top of a regular $1D$ lattice) 
was found capable of sustaining ordered phases~\cite{kim01,medv03}. 
Hence, the inhomogeneous arrangement of particle positions in the $2D$ displacement space, interpreted as a complex network substrate in which 
``spins" undergo ferromagnetic interactions, could arguably be the key ingredient that allows the onset of ordered clusters in the SVM. 

In order to explore this further, let us define the concept of flocks as complex network structures. 
Starting with a disordered initial state in which individuals are 
randomly located within the $L\times L$ displacement space, SVM rules lead to the formation of local structures of 
interacting individuals. These structures, which we call {\it flocks} or {\it clusters}, are not permanent: their 
shape and size evolve with time, with new individuals and sub-flocks merging with them and, conversely, 
other individuals and sub-flocks separating from them.  Moreover, besides the statistical perspective in which flocks are the fundamental building blocks of the ensembles that characterize SVM stationary states, flocks can be regarded as ``domains" that carry key information on the ordering of the system at the mesoscopic level.   

\begin{figure}[t!]
\centerline{{\epsfysize=3.3in\epsfbox{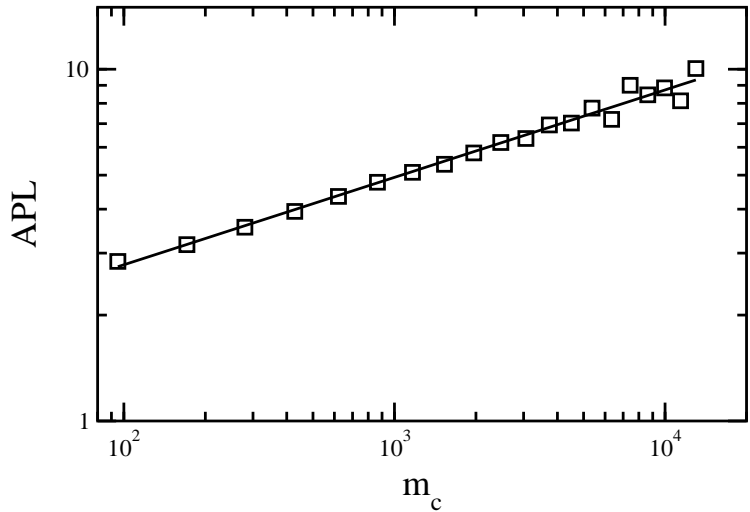}}}
\caption{Log-log plot of the average path length $APL$ as a function of the cluster 
size $m_c$ at criticality. Fits of Eq.(\ref{D_def}) to the data yield $D = 4.0(2)$.}
\label{APL}
\end{figure}

In order to gain insight into the structure of the clusters, let 
us evaluate the average path length, $APL$~\cite{doro08}. 
According to the standard definition used in the study of complex networks, 
for each pair of nodes or individuals $(A,B)$ belonging to the same cluster, the path length $\ell_{AB}$
(also known as chemical distance) is given by the minimum number of links 
that one has to use in order to pass from one node to the other. 
By calculating all the pairwise node-to-node path lengths in the cluster and taking the 
average, one obtains the $APL$, which 
consequently is a characteristic length of the cluster. In Euclidean lattices, the volume of an object is related to its characteristic length by an integer power, 
i.e. the dimension of the object. Based on this observation, as well as on the experience gained in the study of fractal objects, it is customary to define the dimension ($D$) of a complex network according to:
\begin{equation}
APL \propto m_c^{1/D}\ ,
\label{D_def}
\end{equation}                                                      
\noindent where $m_c$ is the complex network size or, in the present context, the cluster mass~\cite{doro08}. 
Figure~\ref{APL} shows a log-log plot of the $APL$ versus $m_c$ for SVM flocks corresponding to the critical point of the second-order phase transition. 
The best fit to the data yields $D = 4.0 (2)$, which strongly suggests that the effective 
dimension of the clusters is $D = 4$. Other complex network measures, not shown here for the sake of space, 
confirm that SVM flocks are characterized by a $4-$dimensional topology~\cite{bagl12}. 

As mentioned above, flocks can be regarded as domains that carry important information on 
the ordering of the system at the mesoscopic level. Therefore, let us now study the onset of order within individual flocks as a 
necessary condition to have macroscopic ordering system-wide. Notice, however, that the detailed 
mechanisms leading to the emergence of global 
order from locally ordered clusters are not well understood yet and remain an open question that lies beyond the scope of the present 
analysis. 

With this aim, here we investigate whether the topology of so-called {\it frozen clusters}, 
once particle displacements and cluster rearrangements are suppressed, is capable by itself of supporting the 
existence of an orientationally ordered phase. In our approach, we first generate configurations of clusters by applying the 
full SVM dynamics. Once the non-equilibrium stationary state is reached, we identify the clusters 
and ``freeze" them, i.e. we disallow any further displacements of the individuals. From that point on, 
the orientation of the particles is allowed to evolve according to Eq.(\ref{anterm}), but subsequent 
displacements (Eq.(\ref{BU})) do not occur. We will refer to this stage as ``restricted SVM dynamics". 
Since the full SVM dynamics has an entanglement between particle 
displacements and XY-type interactions, by resorting to ``frozen clusters" we disentangle these two major components. 
This procedure allows us to explore the relation between local topology and the 
ability for the restricted SVM dynamics to sustain local ordering. 

\begin{figure}[t!]
\centerline{{\epsfysize=3.3in\epsfbox{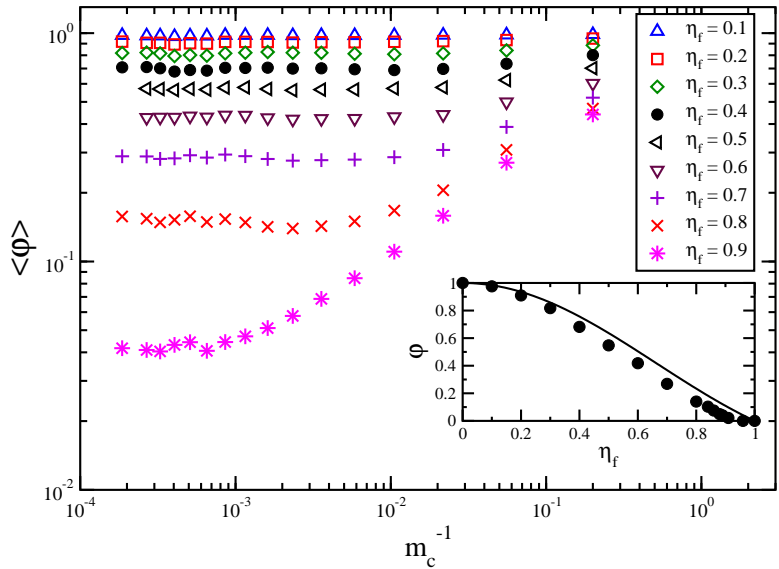}}}
\caption{Log-log plot of the order parameter as a function of the inverse cluster mass for frozen clusters with different noise levels, 
as indicated. Inset: Plot of the asymptotic values of the order parameter, $\varphi_\infty$, versus the noise amplitude $n_f$. 
The solid line shows the mean field (i.e. fully-connected graph) results $\varphi_{MF}= \sin(\pi\eta_f)/\pi\eta_f$~\cite{doss09}.}
\label{AN_OPFrozen}
\end{figure}

Figure~\ref{AN_OPFrozen} shows the dependence of the order parameter $\varphi$ on the inverse cluster mass $m_c^{-1}$. The clusters were first generated using the full SVM dynamics with critical noise. After freezing them, 
the restricted SVM dynamics was applied using different noise values in the $0 < \eta_f < 1$ range, as indicated. 
Even for very large noise amplitudes, the order parameter in the large cluster limit ($m_c\to\infty$) tends to finite values, e.g. $\varphi\simeq 0.04$ for 
$n_f=0.9$. Although we are not presenting here a full finite-size scaling analysis, these results were obtained for a large system size ($N=4\times 10^4$) and 
they provide strong evidence that frozen clusters are capable of sustaining order ($\varphi > 0$) in the thermodynamic limit ($N\to\infty$). 
The Inset to Figure~\ref{AN_OPFrozen} shows the order parameter extrapolations to the $m_c\to\infty$ limit, $\varphi_\infty$, 
as a function of the noise amplitude $n_f$. 
For the sake of comparison, the Inset also shows exact results from the mean field 
solution obtained for an infinite density of individuals~\cite{doss09}, which closely follow the trend of our computer simulation results. 
Recalling that the effective dimension of SVM clusters is $D=4$, we argue that the observed 
mean-field-like behavior is related to the fact that $D = 4$ is the upper-critical dimension of the XY model, which essentially has 
the same symmetries as the SVM defined on frozen clusters.  

Therefore, our findings indicate that the full SVM dynamics creates flocks with enhanced connectivity, which behave as 4-dimensional objects 
compactified into a 2-dimensional displacement space (more details will be published elsewhere~\cite{bagl12}). By virtue of this 
higher-dimensional complex network substrate, the approximately equivalent XY representation (attained by our ``frozen cluster" procedure) 
displays mean-field-like behavior, thus solving the apparent paradox posed by the Mermin-Wagner theorem.  

\section{Conclusions}
In this work, we revisit the Standard Vicsek Model (SVM), a minimal model of self-propelled particles that displays a noise-driven 
phase transition between the ordered phase (in which individuals align with each other and move coherently system-wide) and the 
disordered phase (in 
which the net translational motion averages out to zero). By means of a combined finite-size scaling and dynamical analysis of the SVM, 
we present a full characterization through dynamical and critical exponents and confirm the nature of the SVM phase 
transition as continuous (second-order), in agreement with the early results by Vicsek et al.~\cite{vics95}. 
Finally, we analyze the onset of ordering in SVM flocks by performing a complex 
network analysis and interpreting the results in connection with XY-like spin models. 

In a forthcoming paper~\cite{bagl12}, 
we will extend the complex network analysis of Vicsek flocks to variants of the model, in which perturbations are 
implemented either as Angular Noise (AN) or as Vectorial Noise (VN). 
We believe that this topological approach will provide key insight into the different nature of the ordering phase transitions observed in the Vicsek model under the two noise types, namely continuous second-order for Angular Noise and first-order for Vectorial Noise. Indeed, the connection between
the order of the transition and the topology of the flocks remains an important open
question that certainly deserves further investigation.

\section*{Acknowledgments}
This work was financially supported by  CONICET, UNLP and  ANPCyT (Argentina).

\end{document}